\documentclass{aa}

\usepackage{graphicx}
\usepackage{txfonts}
\usepackage{natbib}

\bibpunct{(}{)}{;}{a}{}{,}
\begin{document}

   \title{A search for radio pulsars in five nearby supernova remnants}

   \author{S.Sett
          \inst{1}
          \and
          R.P.Breton\inst{1} \and C.J.Clark\inst{1} \and M.H. Kerkwijk \inst{2} \and D.L. Kaplan\inst{3}\fnmsep}
   \institute{Jodrell Bank Centre for Astrophysics, School of Physics and Astronomy, The University of Manchester, M13 9PL, UK.\\ 
              \email{susmitasett04@gmail.com}
            \and
             Department of Astronomy $\&$ Astrophysics, University of Toronto, 50 Saint George Street, Toronto, ON, M5S 3H4, Canada. 
         \and
             Department of Physics, University of Wisconsin-Milwaukee, Milwaukee, WI 53211, USA.\\
             }

   \date{received and accepted date}


  \abstract
   {Most neutron stars are expected to be born in supernovae, but only about half of supernova remnants (SNRs) are associated with a compact object. In many cases, a supernova progenitor may have resulted in a black hole. However, there are several possible reasons why true pulsar-SNR associations may have been missed in previous surveys: The pulsar's radio beam may not be oriented towards us; the pulsar may be too faint to be detectable; or there may be an offset in the pulsar position caused by a kick.}
   {Our goal is to find new pulsars in SNRs and explore their possible association with the remnant. The search and selection of the remnants presented in this paper was inspired by the non-detection of any X-ray bright compact objects in these remnants when previously studied.}
   {Five SNRs were searched for radio pulsars with the Green Bank Telescope at 820 MHz with multiple pointings to cover the full spatial extent of the remnants. A periodicity search plus an acceleration search up to 500 m/$\rm s^{2}$ and a single pulse search were performed for each pointing in order to detect potential isolated binary pulsars and single pulses, respectively.}
   {No new pulsars were detected in the survey. However, we were able to re-detect a known pulsar, PSR~J2047+5029, near SNR G89.0+4.7. We were unable to detect the radio-quiet gamma-ray pulsar PSR~J2021+4026, but we do find a flux density limit of 0.08 mJy. Our flux density limits make our survey two to 16 times more sensitive than previous surveys, while also covering the whole spatial extent of the same remnants.}
   {We discuss potential explanations for the non-detection of a pulsar in the studied SNRs and conclude that sensitivity is still the most likely factor responsible for the lack of pulsars in some remnants.}


    \maketitle

\section{Introduction}

The association between supernova remnants (SNRs) and neutron stars was a key prediction leading to the formal identification of pulsars as neutron stars following the discovery of the Crab pulsar \citep{ref:Staelin}. Since then, numerous deep surveys of SNRs for pulsations from young neutron stars have been carried out \citep[e.g.][]{ref:Gorham,ref:VK,ref:Biggs,ref:LLC,ref:1987A,ref:Straal}. The detection and association of pulsars with the remnants has accelerated in the past few years and is a result of high-frequency, targeted searches for radio and gamma-ray pulsars in SNRs \citep{ref:Cam,ref:Gupta,ref:LAT}.

Pulsars associated with SNRs are expected to be young \citep{ref:old} and can be important targets for studying and understanding pulsar properties, such as their braking index \citep{ref:Living}. The measurement of the pulsar braking index is crucial to understanding the underlying pulsar spin-down mechanism. Young pulsars have high spin-down luminosities \citep{ref:spindown} and are more likely to be detected at X-rays and gamma rays, providing observational diagnostics for the rotation-powered neutron star energetics \citep{ref:VK}. For example, \citet{ref:Gelfand} put constraints on the initial spin period of the neutron star in Kes~75 by fitting the observed properties of Kes~75 with the predictions of an evolutionary model of a pulsar wind nebula inside an SNR. Confirmed associations also help in obtaining independent age and distance estimates \citep{ref:age}, which in turn can more accurately constrain the birth properties of neutron stars, namely their period, magnetic field, luminosity, and velocity distributions \citep{ref:Kaplan2}. The detection of a pulsar can also clarify the unusual morphology of a remnant, as in the previous proposed association between PSR B1757$-$ 24 and SNR G5.4$-$ 1.2 \citep{ref:FK}, where the morphology may otherwise lead to misclassification \citep{ref:BH}.  

On the other hand, the non-detection of a pulsar in an SNR may suggest that the supernova resulted in a black hole rather than a neutron star, or that the neutron star received such a large kick upon formation that it is no longer within the region of the SNR \citep{ref:kick}.

However, several selection effects are present when searching for pulsars in SNRs. For example, the choice of observing frequency connects with a number of underlying factors. On the one hand, most pulsars tend to be brighter at low frequencies, which should favour surveys conducted at the lower end of the radio spectrum. However, as most SNRs lie on the Galactic plane and may be located at large distances, effects such as pulse scattering, high dispersion measures, Galactic foreground emission, and emission from the SNRs themselves can hinder such low-frequency surveys \citep{ref:Sanidas}. On the other hand, while higher frequencies may mitigate some of these selection effects, the smaller beam size compared to low frequencies is another factor to consider \citep{ref:Gorham}. Additionally, given the large angular size of some SNRs and the potentially large kick velocities imparted during the supernova explosion, many pointings may be required to survey the full vicinity of the associated remnant. For example, \citet{ref:Schinzel} discovered an association between the SNR CTB~1 and PSR~J0002$+$6216 based on the pulsar's high proper motion and cometary tail, despite a $0.5$~degree offset between the two objects.

We performed a sensitive survey of five nearby SNRs with the 100-m Robert C. Byrd Green Bank Telescope (GBT) in Green Bank, West Virginia, USA, to search for new neutron stars by their radio pulsations. The SNRs chosen for this survey are all a result of core collapse supernovae (cc-SNe) and are expected to produce a compact object after the explosion. Several similar SNRs were searched for X-ray bright compact objects by \citet{ref:Kaplan3}. However, no X-ray bright compact sources were detected in the remnants. This survey attempts to search a subset of the remnants studied by \citet{ref:Kaplan1} and detect potential neutron stars through their radio pulses. In this paper, we report our results of the survey of five SNRs.
   \begin{table*}
      \centering
      
         \label{Table:1}
     $$
         \begin{tabular}{|p{1.5cm}p{0.5cm}p{0.5cm}p{0.9cm}p{1.4cm}p{1.5cm}p{0.8cm}p{1.0cm}p{1.5cm}p{1.7cm}p{1.2cm}|}
         \hline
SNR        & \begin{tabular}[r]{@{}l@{}}Dist\\ (kpc)\end{tabular} & \begin{tabular}[r]{@{}l@{}}Size\\  (')\end{tabular} & \begin{tabular}[r]{@{}l@{}}Age\\($10^{3}$ yr)\end{tabular} & \begin{tabular}[r]{@{}l@{}}Number of \\ pointings\end{tabular} & \begin{tabular}[r]{@{}l@{}}Total\\observation\\time (hr)\end{tabular} & \begin{tabular}[r]{@{}l@{}}$\rm T_{\rm sky}$\\ (K)\end{tabular} & \begin{tabular}[r]{@{}l@{}}$\rm S_{\rm SNR}$\\  (Jy/ \\ beam)\end{tabular} &   \begin{tabular}[r]{@{}l@{}}Average \\ $\rm S_{min}$ \\(mJy)\end{tabular} & 
\begin{tabular}[c]{@{}l@{}}Required \\ 2D velocity\\(km/s)\end{tabular} &  \begin{tabular}[c]{@{}l@{}}Previous \\ $\rm S_{min}$ \\(mJy)\\ \end{tabular}
 
        \\ 
            \hline
            
G53.6$-$2.2 & $2.8^{a}$ & 33 & $7^{b}$ & 2 & 4.5 & 14.11 & 1.92 & 0.11 & 1900 & 0.2\\
G78.2+2.1 & $1.5^{c}$ & 60 & $7^{d}$ & 5 & 6.8 & 39.56 & 22.13 & 0.15 & 1800 & 2.4\\ 
G89.0+4.7  & $1.7^{e}$ & 105  & $16^{f}$ & 10 & 7.0 & 16.41 & 3.71 & 0.14 & 1600 & 1.0\\ 
G116.9+0.2 & $1.6^{g}$ & 34  & $7^{h}$ & 7 & 2.3 & 10.58 & 1.76 & 0.19 & 1100 & 0.8 \\ 
G156.2+5.7 & $1.3^{i}$ & 110 & $15^{j}$ & 27 & 7.3 & 7.31 & 0.10 & 0.16 & 1300 & 0.7 \\

            \hline
            
         \end{tabular}
     $$ 
     \caption[]{Supernova remnants targeted in this survey without compact objects and with distances less than 3 kpc. Ages are from Sedov-phase approximations using X-ray temperatures. The size is the diameter of the remnants. The total observation time is the time each remnant has been observed. $\rm T_{\rm sky}$ is the sky temperature calculated from \citet{ref:Has} assuming a spectral index of $-$2.6 and an 820 MHz observing frequency. The $\rm S_{\rm SNR}$ is the flux of the SNR per pointing calculated using data from \citet{ref:Green}. Also shown are the average flux density thresholds of the remnants, the required velocity of the pulsar to escape the field of view studied in the survey, and the previous recorded minimum flux densities. All the SNRs except G53.6$-$2.2 were observed by \citet{ref:LLC} down to the minimum flux densities given in the column 'Previous $\rm S_{\rm min}$'. SNR G53.6$-$2.2 was observed by \citep{ref:Gorham} down to a sensitivity limit of 0.2 mJy. References are: a) \citet{ref:G}, b) \citet{ref:Goss}, c) \citet{ref:Land}, d) \citet{ref:LGR}, e) \citet{ref:Byun}, f) \citet{ref:Laz}, g) \citet{ref:Yar}, h) \citet{ref:Hail}, i) \citet{ref:Pff}, and j) \citet{ref:Borow}.}
   \end{table*}

\section{Targeted supernova remnants}
\label{sec:details}
In this section, we discuss the important properties and previous pulsar searches of the five remnants targeted in our survey. Table \ref{Table:1} provides the distance, size, and age of the SNRs. These SNRs were chosen due to a suitable combination of relatively small distance (less than 3 kpc), low inferred age (less than 20000 years), small angular extent, and low sky and remnant fluxes. These factors imply that full coverage of the remnants using a reasonable number of pointings and deep radio luminosity limits can be achieved for all of them.

$\mathbf{SNR~G53.6-2.2}$ (also known as 3C~400.2) is a mixed-morphology SNR, that is, it consists of a radio shell as well as centrally brightened X-ray emission \citep{ref:Bro}. G53.6$-$2.2 was searched for pulsars by \citet{ref:Gorham} using the Arecibo 305-m telescope at 430 MHz. They used long-duration observations (20 minutes to 2 hr) with rapid sampling to reach a sensitivity of 0.2 mJy. However, no pulsars were detected in the survey. 

$\mathbf{SNR~G78.2+2.1}$ (also known as $\gamma$ Cygni) is a shell-type SNR that has been imaged in radio waves to gamma rays. The SNR is located in the Cygnus X star-forming region \citep{ref:LGR}. Radio observations of G78.2+2.1 show that the radio diameter of the remnant is approximately 60 arcmin \citep{ref:H}. A gamma-ray source, 2CG~078+2, was discovered in the field of the remnant with the COS B satellite \citep{ref:Swa}. This unidentified source was suspected to be a pulsar or to be the interactions of accelerated energetic particles with matter and radiation \citep{ref:Bykov}. Around the time of acquiring our observations, a blind search with the \textit{Fermi} Large Area Telescope (LAT) established the unidentified source as a radio-quiet gamma-ray pulsar, PSR~J2021+4026 \citep{ref:LAT}. Measurement of the neutral hydrogen column density using the X-ray spectrum of PSR~J2021+4026 is consistent with that of the diffused emission located in the central and south-eastern part of the SNR \citep{ref:Hu}. It was also noted that these values agree with the neutral hydrogen column density inferred from the HI radio absorption spectrum \citep{ref:LGR}. These results indicate that the pulsar emission, diffuse X-ray emission, and the radio shell are at the same distance, and hence they support the association of the pulsar with the remnant. Targeted searches for the radio counterpart of the pulsar have not yielded a positive result, and hence the pulsar is considered to be radio-quiet \citep{ref:Hui}. Our non-detection confirms this (see Section~\ref{sec:results}).

$\mathbf{SNR~G89.0+4.7}$ (also known as HB~21) is a large, mixed-morphology SNR. It was discovered by \citet{ref:Han} at 159 MHz. A pulsar, PSR~J2047+5029, was detected by \citet{ref:Janssen} with the Westerbork Synthesis Radio Telescope at a frequency of 328 MHz. However, the distance estimates of the pulsar (1.7 kpc, combining all distance estimates from \citealt{ref:Byun}) and the SNR (4.4 kpc, \citealt{ref:CL}) differ by a factor of three. The characteristic age of the pulsar, 1.7 Myr, is also two orders of magnitude greater than the age of the SNR (16 kyr). If we require the age of the pulsar to be consistent with that of the SNR, its birth spin period would be similar to the current one of 0.445s. This is extremely slow when compared to the typically estimated pulsar birth periods \citep{ref:Fauche}, suggesting that the pulsar has to be far older than the estimated SNR age. Due to the above reasons, the pulsar is not believed to be associated with the remnant \citep{ref:Janssen}. Further searches for an associated radio pulsar by \citet{ref:Biggs} and \citet{ref:LLC}, with flux density limits of 13 mJy (400 MHz) and 1 mJy (606 MHz), respectively, were unsuccessful. However, it should be pointed out that a handful of central compact objects (CCOs) with clear associations with SNRs dubbed `anti-magnetars' have been discovered to be young, but slowly rotating pulsars \citep{ref:anti}. A measurement of the proper motion of PSR~J2047+5029 would help clarify its possible association.

$\mathbf{SNR~G116.9+0.2}$ \citep[also known as CTB~1,][]{ref:Wilson} is an oxygen-rich, mixed-morphology SNR. It has a complete shell in both optical and radio. The uniform optical and radio shells that define CTB~1 are indicative of a blast wave extending into a relatively uniform interstellar medium \citep{ref:Laz}, which may provide a kick velocity to the compact object produced in the explosion. A radio pulsar survey by \citet{ref:LLC} did not yield a positive result. However, a gamma-ray pulsar, PSR~J0002+6216 \citep{ref:Clark}, was recently detected by the Einstein@Home survey of unidentified \textit{Fermi}-LAT sources. \citet{ref:Zyuzin} suggested an association with the remnant due to the consistency between the distance of the pulsar and the SNR. This association has recently been confirmed by \citet{ref:Schinzel}, who measured a high proper motion and a long bow-shock pulsar wind nebula that both point away from the SNR. While the pulsar is radio-loud and has been observed in the S and L bands by the Effelsberg Telescope \citep{ref:Colin}, we were unable to detect this pulsar in our survey as it is out of our field of view (see also Section~\ref{sub:reasons}).

$\mathbf{SNR~G156.2+5.7}$ (also known as RX04591+5147) was initially discovered in X-ray with an X-ray astronomy satellite, ROSAT \citep{ref:Pff}. G156.2+5.7 has a spherical shell and is one of the brightest SNRs in X-rays \citep{ref:Pff}. \citet{ref:LLC} searched the remnant with the 76-m Lovell telescope, but no pulsars were detected. There are no compact objects associated with the remnant to date.

\section{Observation and data reduction}
The five SNRs discussed above were observed with the GBT (Proposal ID:GBT/10B--044). The survey was conducted with the Prime Focus (PF1) receiver, set to the 680--920 MHz frequency band with the 200 MHz bandwidth intermediate frequency(IF) filter mode, feeding the Green Bank Ultimate Pulsar Processing Instrument(GUPPI) back end \citep{ref:GUPPI}. The field of view of the receiver is 12.5 arcmin. The whole spatial extent of the remnants was observed with this configuration to account for the possible offset position of the compact object due to large kick velocities. The chosen configuration has a large beam size and hence minimises the pointings required for the complete survey. It also arguably provides a balance between background sky temperature and signal, given that pulsars usually have shallower spectral indices than the sky background \citep[i.e $-1.4$ vs. $-2.6$, respectively,][]{ref:Bates,ref:Has}. Table \ref{Table:1} shows the total number of pointings required to cover the whole spatial extent of the SNRs and the total observation time for each remnant.

The data were processed with PRESTO \citep{ref:Ransom}. The sampling time was 61.44\,$\mu$s, and the number of channels was 2048. We used 128 subbands in order to strike a balance between computational efficiency and survey sensitivity. The de-dispersion plan was created using PRESTO's \emph{ddplan} routine. While the NE2001 model \citep{ref:CL} predicts a maximum dispersion measure(DM) of 500 pc $\rm cm^{-3}$ for all the remnants in this survey, we chose to search DMs up to 2000~pc~cm$^{-3}$, in steps of 0.03~pc~cm$^{-3}$ below $\textrm{DM} = 300$~pc~cm$^{-3}$ and in steps of 0.05~pc~cm$^{-3}$ above, to account for potential extra contributions from the remnant surroundings. The de-dispersed data was then fast Fourier transformed to search for periodicity.

An acceleration search was also performed to search for binary pulsars. The maximum acceleration of a binary system with orbital period $P$ is
\begin{equation}
z = \left(\frac{G m_{c}^{3}}{m_{\rm tot}^{2}}\right)^{1/3}\left( \frac{2\pi}{P}\right)^{4/3} \,,
\end{equation}
where $m_c$ is the mass of the companion and $m_{\rm tot}$ is the total mass of the system. A pulsar with spin frequency $f$ experiencing a constant acceleration has an apparent spin-down rate of $\dot{f} = f z/c$. For an integration time of $T_{\rm obs}$, this acceleration range must be searched with step size $\Delta \dot{f} = 1/T_{\rm obs}^2$, while acceleration searches lose sensitivity if $T_{\rm obs} \gtrsim P/10$ \citep{ref:Ransom}. Assuming a canonical pulsar mass of $1.4\,M_{\odot}$, we therefore designed our search to be sensitive to binaries with companions lighter than a neutron star ($m_{\rm c} < 1.4\,M_{\odot}$), orbital periods at least five times longer than the integration times, and spin frequencies below $f = 100$\,Hz. For our shortest (longest) pointings, this range corresponds to $z=\pm506$~m~s$^{-2}$ ($z=\pm26$~m~s$^{-2}$) and $n=\pm160$ ($n=\pm704$) acceleration steps.

We selected candidates from the acceleration search that had a PRESTO-reported significance above $6\sigma$ and a signal-to-noise ratio above $5$ and removed duplicate and harmonically related candidates. Remaining candidate signals were folded and visually inspected to classify them as radio interference or promising pulsar candidates.

We estimated the sensitivity of our survey by applying the pulsar version of the radiometer equation to find the limiting flux density given by \citep{ref:Lor},
\begin{equation}
S_{\rm min} = \frac{\beta \left(S/N\right)_{\rm min} \sqrt{D}}{\sqrt{nBT_{\rm obs}}}\left ( \frac{T_{\rm sys}+ T_{\rm sky}}{G}+S_{\rm snr} \right ) \,,
\label{equn:min}
\end{equation}
where $\beta = 1.5$ is a predetermined factor due to losses and system imperfections and $\left(S/N\right)_{\rm min} = 5$ is the minimum signal to noise at which the pulsar is expected to be detected. With our observing setup, the number of polarisations was $n = 2$, the instrument bandwidth $B = 200$\,MHz, the system temperature $T_{\rm sys} = 29$\,K, and gain $G = 2$\,K~Jy$^{-1}$. The latter two were assumed as per GBT specifications\footnote{\url{https://science.nrao.edu/facilities/gbt/proposing/GBTpg.pdf}}. The sky temperature, $T_{\rm sky}$, was calculated from \citet{ref:Has} assuming a power law spectrum with an exponent $-2.6$ and 820 MHz observing frequency. The $S_{\rm SNR}$ is the flux density of the SNR per beam and was calculated using data from \citet{ref:Green}. Finally, we assumed a pulse duty cycle $D = 0.05$ and used the appropriate integration time of each pointing, $T_{\rm obs}$.

The de-dispersed time series were also searched for single pulses using PRESTO's \textit{single\_pulse\_search.py} python routine \citep{ref:Ransom}. No excesses of significant candidate pulses were detected in the searches towards any SNR.

\section{Results}
\label{sec:results}
No new pulsars were detected in this survey. However, we were able to blindly re-detect PSR~J2047+5029 in SNR G89.0+4.7, at a DM of 107.104 pc\,$\rm cm^{-3}$, with a significance of $11.8\sigma$, a S/N of 6.5, and an estimated flux density of $0.2$~mJy. As discussed in Section \ref{sec:details}, this pulsar is not believed to be associated with the SNR. The flux density of this pulsar estimated by \citet{ref:Janssen} is 2.5 mJy at a central frequency of 328 MHz. If we assume a power law spectrum $S(f) \propto f^{\alpha}$, our detection yields a spectral index $\alpha = -1.9$, which is compatible with the typical range for pulsars, $-3 < \alpha < -1.3$ \citep{ref:Johnston}.

We were unable to detect the radio-quiet pulsar PSR~J2021+4026, either blindly or by folding using the gamma-ray timing ephemeris \citep{ref:second}. Using Equation~(\ref{equn:min}), we find a flux density limit of 0.08 mJy at a S/N of 6, which is an order of magnitude better than the survey by \citet{ref:Tre}. PSR~J0002+6216, the gamma-ray, radio-faint pulsar associated with SNR G116.9+0.2, was also not detected due to its location outside the SNR and our beams (see Section~\ref{sec:details}).

We calculated the upper limits on the flux density of the survey for each SNR using Equation \ref{equn:min}. We also calculated the 2D minimum velocity required for a pulsar to travel from the centre of the SNR to the approximate edge of the surveyed region (which provides slightly more coverage than the SNR extent) in a time corresponding to the estimated age of the remnant reported in Table \ref{Table:1}. These limits are also reported in Table \ref{Table:1}.

\section{Discussion}\label{sub:reasons}
The average flux density limits of our survey can be compared with the previously reported limits given in Table \ref{Table:1} and discussed for each individual SNR in Section \ref{sec:details}. Overall, we achieved sensitivity limits between $\sim$2 and 16 times deeper than previous surveys. Furthermore, these past surveys only cover the central regions of the SNRs where the pulsars are expected to be born. In comparison, we surveyed the whole spatial extent of the SNRs.

Despite improving on the flux density limit, it is likely that low radio luminosity is a primary factor accounting for the non-detection of pulsars in the empty SNRs. Not all pulsars of the pulsar population could be detected in our survey. For example, the young gamma-ray pulsars ~J0106+4855 and ~J1907+0602 are extremely faint in radio, their radio flux density being around 3 $\mu$Jy \citep{ref:second}. We also computed pseudo-luminosity upper limits ($L_{\rm min} = S_{\rm min}~ d^{2}$) using the values of $d$ and $S_{\rm min}$ provided in Table~\ref{Table:1}. We compared these to the luminosities of the known pulsars in the Australia National Telescope Facility(ATNF) catalogue\footnote{\url{https://www.atnf.csiro.au/research/pulsar/psrcat/}} \citep{ref:ATNF}, extrapolating the $400\,$MHz and $1400$\,MHz luminosities reported therein to our $800$\,MHz observing frequency assuming a spectral index of $-2$. About 10\% of the pulsars (in accordance with our most sensitive pointing) that have reported luminosities at 400 MHz and 1400 MHz and are not a part of a binary system would not have been detected by our survey. Since the low-luminosity pulsar population is severely under-sampled (as they are by nature harder to discover), we conclude that it is quite likely that these SNRs contain pulsars that are simply too faint to be detected by our survey.

There are other possibilities that could account for the lack of a neutron star in an SNR. For instance, it is possible that a pulsar lies within the remnant but that the radiation is not beamed towards the observer. The typical beaming fraction is assumed to be $\approx$ 20 \% in the radio band \citep{ref:m}. A larger beaming fraction for the young population of pulsars, as suggested by \citet{ref:Ravi}, would make it more difficult to reconcile with our results. Observations show that pulsar beams can be patchy, and therefore the pulsar could remain undetectable despite having a large beam. However, in this case, it may be possible that the pulsar is visible in gamma rays due the gamma-ray beam typically covering a larger range of latitudes. This is the accepted explanation for the non-detection of the radio-quiet gamma-ray pulsar PSR~J2021+4026 in SNR ~G78.2+2.1. The other gamma-ray pulsar, PSR~J0002+6216 in SNR~ G116.9+0.2, was not detected here due to its large angular distance from the SNR, which we discuss later. We searched the \textit{Fermi}-LAT fourth source catalogue \citep{ref:4FGL} for any unidentified gamma-ray source within a radius of one degree from each of the five SNRs studied in our survey, but we were unable to find any such source that could be classified as a pulsar other than the two already known gamma-ray pulsar associations.

One of the other possibilities is that the pulsar's magnetic field may take a considerable amount of time to develop. If the growth timescale is $10^{5}$ years or more, then even a rapidly spinning neutron star could still be undetectable \citep{ref:Bonanno,ref:BR}. In this case, even if a pulsar were present in the remnant, it would not be emitting radio waves and should be detected as a CCO. Good examples supporting such a scenario are the young CCOs RX J0822-4300 in SNR Puppis A \citep{ref:Got} and the faint CCO in SNR Cas A \citep{ref:Tan}. However, these neutron stars are still hot from their birth and so should be detectable in X-rays. Of the SNRs searched in this survey that remain unassociated with a pulsar, SNR ~G156.2+5.7 has been searched in X-ray for compact objects down to a limit of $10^{32}$ ergs $s^{-1}$ \citep{ref:Kaplan1}. However, no obvious X-ray sources that could be neutron stars were detected. In order for neutron stars present in these SNRs to be too faint to be observed in X-rays, they would require a cooling process that differs from the predicted cooling processes of young pulsars, such as those in Vela \citep{ref:Page} and 3C~ 58 \citep{ref:Slane}. For example, \citet{ref:Kaplan1} suggest that if a neutron star is massive enough to support direct Urca \citep[beta decay and electron capture,][]{ref:DURCA}, then the appearance of superfluidity as it cools would lead to a powerful neutrino emission that accelerates the cooling, allowing it to become invisible within decades.

Another plausible explanation is that the neutron star formed in the supernova explosion has undergone a large velocity kick and is no longer in our viewing field. Such kicks may make it hard to associate the pulsar with a nearby SNR \citep{ref:Lai}. The minimum kick velocities that pulsars in the SNRs in this survey would have had to experience to escape our surveyed regions are given in Table \ref{Table:1}. They all exceed 1000 km/s. \citet{ref:Fauche} predicted that most pulsars should have kick velocities slower than 400 km/s, which is well below the velocities given in Table~\ref{Table:1}. However, extreme cases are known, such as PSR~B2011+38 with a velocity of $\sim 1600$\,km/s \citep{ref:B2011}, as well as several others with more poorly constrained distances \citep[e.g. PSR~B2224+65 in the Guitar Nebula, 1640\,km/s;][]{ref:Velocity}. A similar explanation accounts for the non-detection of PSR~J0002+6216 in SNR ~G116.9+0.2, which was recently found to be travelling at about 1100 km/s away from the remnant's centre \citep{ref:Schinzel}. This is thought to be due to the hydrodynamic instabilities in the supernova explosion.

Even though most of the remnants studied are transparent at radio wavelengths, it is possible that the immediate environment of the central star has a relatively high gas density, which would cause unusually large scattering and absorption \citep{ref:Zanardo}. An example of such scattering is that of the Crab Pulsar in the Crab Nebula \citep{ref:Laura}. More such extreme events could heavily hinder the chances of detecting a pulsar.

Finally, it is possible that the supernova explosion resulted in a black hole instead of a pulsar. It is expected that between 13 \% and 25 \% of cc-SNe produce a black hole \citep{ref:Heger}. The catalogue of galactic SNRs \citep{ref:Green2019} has 294 SNRs. Of the 294 SNRs, about 75 \% are believed to be the result of cc-SNe \citep{ref:Cap}. About half of these have a possible neutron star, pulsar, or pulsar wind nebula association. As none of the three remaining unassociated SNRs that we studied have an associated pulsar wind nebula, it is possible that they could have formed a black hole instead of a neutron star.

\section{Conclusion}
We have performed a sensitive search for young pulsars in five nearby SNRs. No new pulsars were discovered in this survey, although two out of the five SNRs are now known to be associated with pulsars. One of these, PSR~J2021+4026, is gamma-ray-loud but thought to be radio-quiet. The other one, PSR~J0002+6216, is both gamma-ray-loud and radio-loud but located outside the SNR due to an extremely high kick velocity. We were able to re-detect PSR~J2047+5029 in SNR~G89.0+4.7, but this pulsar is not believed to be associated with this SNR.

We obtained improved sensitivity limits for all five SNRs. Our work adds to the growing evidence that the likely dominant factor responsible for the lack of pulsar-SNR association is limited survey sensitivity. Higher sensitivity surveys from next-generation facilities, such as MeerKAT, the Five Hundred Metre Aperture Radio Telescope (FAST), and the Square Kilometre Array (SKA), will likely uncover many new pulsar-SNR associations. However, it is also now apparent that a fair fraction of pulsars will lie outside the SNR due to high kick velocities and therefore require surveys to search an extended region beyond their boundaries. We can also infer that some of the SNRs may not host a neutron star but rather a black hole, and that some of them may be invisible due to propagation and beaming effects.

\begin{acknowledgements} 
R.P.B. and C.J.C. acknowledge support from the ERC under the European Union's Horizon 2020 research and innovation programme (grant agreement No. 715051; Spiders). We would like to thank the GBT for the observations. The Green bank Observatory is a facility of the National Science Foundation operated under cooperative agreement by Associated Universities, Inc. 
\end{acknowledgements}


\bibliographystyle{aa}
\bibliography{main.bib}
\end{document}